\documentclass[12pt,preprint2]{aastex}

\newcommand\plottwovert[2]{\centering \leavevmode
 \includegraphics[width={0.99\columnwidth}]{#1} \hfil
 \includegraphics[width={0.99\columnwidth}]{#2}}

\begin{document}

\title{The effect of mixing on metallicity gradients in the ICM}

\author{M.Br\"{u}ggen\altaffilmark{1,2}}
\altaffiltext{1}{International University Bremen, Campus Ring 1, 28759 Bremen, Germany}
\altaffiltext{2}{Max-Planck Institut f\"ur Astrophysik, Karl-Schwarzschild-Str 1, 85740 Garching, Germany}

\begin{abstract}
It is generally argued that most clusters of galaxies host cooling
flows in which radiative cooling in the centre causes a slow
inflow. However, recent observations by Chandra and XMM conflict with
the predicted cooling flow rates. It has been suggested that radio
jets that are situated at the centre of clusters of galaxies can
assist in reducing the deposition of cold gas. Radio jets inflate
cavities of hot radio plasma that rise through the cluster atmosphere
and thus stir the intra-cluster medium. In this letter it is investigated
whether this scenario is consistent with the pronounced metallicity
gradients that have been observed in the cores of clusters.
\end{abstract}

\keywords{galaxies: active - galaxies: clusters:
cooling flows - X-rays: galaxies}

\label{firstpage}

\sloppypar

\section{Introduction}

The X-ray surface brightness of many clusters of galaxies shows a
strong central peak which is generally interpreted as the signature of
a cooling flow (Cowie \& Binney 1977, Fabian \& Nulsen 1977, Sarazin
1988, Fabian 1994). However, the simple cooling flow model conflicts
with a growing number of observations that show that while the
temperature is declining in the central region, gas with a temperature
below $\sim$1 keV is significantly less abundant than predicted.
Moreover, the level of star formation found in present-day cooling
flows is insufficient to explain the mass deposition rates inferred
from X-ray observations (B\"ohringer et al. 2001). 

Many clusters of galaxies host a radio source at their centres. Around
71\% of all cD galaxies in cooling flows show evidence for radio
activity which implies that either the sources are active for longer
than 1 Gyr or that they recur with high frequency (Burns 1990).  These
radio sources emit large quantities of hot relativistic plasma into
the intracluster medium (ICM). The displacement of thermal plasma by
the relativistic plasma can create `holes' in the X-ray surface
brightness such as observed in the Perseus cluster (B\"ohringer et
al. 1993). Holes in the X-ray emission are also seen in the radio
lobes of the Hydra A cluster (McNamara et al. 2000), Centaurus A
(Saxton, Sutherland and Bicknell 2001) and other clusters (Fabian et
al. 2000, Mazzotta et al. 2001). It has also been shown that bubbles
are difficult to detect and that a substantial amount of energy may be
hidden in this form in clusters (Br\"uggen et al. 2002). \\

The radio jets inflate pockets of low-density radio plasma that rise
buoyantly through the cluster atmosphere. In the course of this rise
the cavities become Rayleigh-Taylor unstable and form plumes and
mushroom-like structures (Churazov et al. 2001, Br\"uggen \& Kaiser
2001). It has been shown that the rising bubbles can lead to mixing and a
significant restructuring of the inner regions of a cooling flow
(Churazov et al. 2001, Br\"uggen \& Kaiser 2001, Nulsen et
al. 2001). This process can reduce the deposition of cool gas without
directly introducing heat and thus can form an effective machanism to
couple the mechanical power of the radio jets to the cooling gas. In
some cases this may be sufficient to quench the cooling flow (Binney
2001, Binney \& Tabor 1995).\\

However, pronounced metallicity gradients have been detected in the
core of some clusters. Chandra observations of Hydra A have shown
a factor 2 increase in the Fe and Si abundances within the central 100
kpc (David et al. 2001). Fukazawa et al. (2000) measured metal
abundances in the ICM of 34 clusters using ASCA data and also found
increments in Si and Fe abundances around cD galaxies.  Similar
results have been found by Ezawa et al. (1997), Ikebe et al. (1997),
Finoguenov \& Ponman (1999) and Finoguenov et al. (2000). Fukazawa et
al. (2000) found that the central Fe abundances of cD type clusters
are by up to a factor of 3 higher than their spatially averaged
values, while non-cD type clusters do not show this Fe excess at the
centre. This indicates that the excess metals are probably ejected
from the central cD galaxy but the exact origin of these metals is
still subject to debate. However, the metals have to be produced at
the cluster centre because gravitational settling of the heavy ions
takes much longer than the age of the Universe (Sarazin 1988). (The
abundance gradients occur mostly around cD galaxies which are believed
to sit still at the bottom of the cluster potential. Therefore, it
seems unlikely that the metals have been stripped off by the ICM.)

The presence of steep metallicity gradients constrains the amount
of mixing that can take place in a cluster. If stirring motions induced by
buoyant bubbles erase the metallicity gradient faster than the metals
can be replenished at the centre, it will be difficult to explain the
observed metallicity gradients. In order to calculate the effect of
rising bubbles on the metallicity gradients of a cluster, I performed
three-dimensional hydrodynamical simulations and employed tracer particles
to compute the metallicity gradients.

\section{Method}

The simulations were obtained using the parallel version of the
ZEUS-3D code which was developed especially for problems in
astrophysical hydrodynamics. The code uses finite differencing on an
Eulerian grid and is fully explicit in time. For a detailed
description of the algorithms and their numerical implementation see
Stone \& Norman (1992a, b). In order to study the mixing of the ICM,
the ZEUS code was modified to follow the motion of tracer
particles. These tracer particles are passively advected with the
fluid and their positions are written to a file at regular
intervals.\\

In our simulations we employed an ideal gas equation of state and we
ignored the effects of magnetic fields and rotation.  The cooling time
in our cluster model is of the order of 500 Myrs which is sufficiently
long compared to the time scales considered here so that we ignored
it. Moreover, numerical experiments have shown that the neglect of
cooling does not influence the large-scale dynamics of the fluid over
the time-scales considered here. The simulations were computed on a
Cartesian grid in three dimensions.  The serial version of the code
was run on a SUN ULTRA workstation and the parallel version on an SGI
ORIGIN 3000. The computational domain spanned 30 kpc in height with a
base area of 10 kpc $\times$ 10 kpc. This domain was covered by 450
$\times$ 150 $\times$ 150 equally spaced grid points. On the
boundaries outflow conditions were chosen.\\

The initial mass and temperature distributions were modelled on the
Virgo Cluster as given by Nulsen \& B\"ohringer (1995). The gas
density distribution was found by assuming hydrostatic equilibrium to
maintain an initially static model. Here the results from two runs are
reported: run 1 and run 2. At each timestep hot gas was injected into
a spherical region which was situated at a distance of $d=9$ kpc from
the gravitational centre. This spherical region had a radius of
$r_{\rm b}=0.7$ kpc and $r_{\rm b}=1.$ kpc for run 1 and 2,
respectively. The injected gas had zero initial velocity and the
energy injection rates were $L=4.4\times 10^{41}$ erg s$^{-1}$ (run 1)
and $L=1.3\times 10^{43}$ erg s$^{-1}$ (run 2). This particular set of
parameters was chosen because they could reproduce a lot of
observational features of late-stage radio activity. As shown
elsewhere (Br\"uggen et al. 2002) the latter luminosity is close to
the maximum rate of thermal energy injection that produces a buoyant
plume. Even larger energy injection rates have the effect of
evacuating the core before the bubble can start to rise. It was
assumed that the material in the injection region is supplied by the
jet of an Active Galactic Nucleus (AGN). The flow from the centre of
the source to the injection region is taken to be ballistic while the
injection region itself may be identified with the location of a
strong internal shock that brings the jet into pressure equilibrium
with the surrounding gas. This picture motivates the choice of
parameters for the position of the injection region for which $d=9$
kpc is a reasonable choice. We have also explored slightly different
parameters in Br\"uggen et al. (2002) and found that the resulting
flow is not strongly dependent on the choice of these parameters.

At the start of the simulation 24,000 tracer particles were
distributed uniformly throughout the computational domain
(Fig. 1). The gas was treated as a single fluid and assumed to obey a
polytropic equation of state with $\gamma = 5/3$. For more details of
the simulation the reader is referred to the paper by Br\"uggen et
al. (2002).\\

Finally, I should address some issues related to the accuracy of
these kinds of finite-difference hydrodynamical simulations. While the
code can simulate large-scale mixing due to Rayleigh-Taylor and
Kelvin-Helmholtz instabilities, it does not include real diffusion of
particles. Any observed diffusion is therefore entirely
numerical. Moreover, numerical viscosity is responsible for
suppressing small-scale instabilities at the interface between the
bubble and the cooler, surrounding, X-ray emitting gas.  To assess the
effects of numerical viscosity, I have repeated some simulations on
grids with lower resolution. From these experiments I can conclude that
'global parameters' such as the position and size of the 'mushrooms'
are relatively insensitive to the resolution. However, the detailed
morphology on small scales does depend on the resolution and the
choice of initial conditions.

\section{Results}

Fig.\ 1 shows the positions of the tracer particles at the start of
the simulation and after a simulated time of 63 Myrs (run 2). The
tracer particles are color coded according to their initial positions
as shown in the left panel. One can see how the bubble displaces the
material and lifts the tracer particles up from the cluster centre. In
Fig.\ 2 the differences between the original iron abundance profiles
and that at a time $t$ after the start of the energy injection are
shown. I assumed an initial iron abundance given by $A_{\rm
Fe}=0.703-0.0053\cdot (r/{\rm kpc})$, where $A$ is measured in units
of the solar abundance and $r$ denotes the distance from the centre of
the cluster measured in kpc. This simple fit is based on the
observations of Hydra A by David et al. (2000). The abundances have
been averaged over horizontal slices through the computational
volume. One should note that in computing the abundances I included
only those metals of the ICM that already were present at the start of
the simulation. I ignore those metals that may have been injected with
the hot gas. Therefore, the changes in the abundances are caused
solely by the displacement and mixing of the ICM. Thus, one obtains an
upper limit on the abundance changes because the injected gas may
contain metals, too. Furthermore, one can note that the bubbles might
dredge up material from the central regions of the active galaxy which
is likely to be enriched by stars of supersolar abundances.\\

Fig.\ 3 shows the same as Fig.\ 2 only that the abundances have been
averaged over semispherical shells. This way the observations are
simulated where the spectra are averaged over concentric rings about
the cluster centre. Here I average over semi-spherical shells assuming
that the jet is part of an ambipolar jet. Now, the abundance changes
are much smaller at larger radii due to the larger volume to average
over. The only noticeable decrease occurs within the inner 5 kpc. For
run 1 the abundance changes found are smaller than the accuracy of the
observed values which is of order of $\pm 0.1$ solar (David et
al. 2001). For run 2, the metallicity at the inner edge of the core is
decreased by nearly 0.25 solar but averaged over the inner 5 kpc the
decrease is roughly 0.1 solar.

\section{Conclusion}

It was found that mixing by buoyant bubbles has a relatively weak
impact on the metallicity gradients found in clusters. The simulations
were stopped after 120 Myrs but during this time supernovae in the cD
galaxy will have replenished some of these metals.

The fact that the Si/Fe abundance ratio of the central excess metals
is close to the solar value suggests that typa Ia supernovae are the
main suppliers for these metals. Over the course of our simulation the
Fe abundance within the central 5 kpc was decreased on average by
$\Delta A \sim 0.1$ solar (run 2). Given that the mass within the same
radius is $M_{\rm 5 kpc} \sim 10^{10} M_{\odot}$ this corresponds to
an Fe mass of $\Delta M_{\rm Fe}\sim 10^4 M_{\odot}$. Adopting
the supernovae yields that have been averaged over a Salpeter IMF as
given by Finoguenov et al. (2000), $y_{\rm Fe}=0.74 M_{\odot}$ (type
Ia), it requires less than $N_{\rm SN}=\Delta M_{\rm Fe}/y_{\rm Fe} \sim
2\times 10^4$ supernovae to replenish the dredged-up mass of
iron. Over the time of the simulation this would correspond to a
supernova rate of $\sim 2\times 10^{-4}$ yr$^{-1}$ which is still
less than the expected type Ia supernova rate for a typical central
galaxy (van den Bergh 1987). Hence, the metals lost through the mixing
by bubbles can be replenished by supernovae from the old stellar
population in the central galaxy.

Therefore, one can conclude that the heating scenario in cooling flows is
consistent with the measured metallicity gradients.

Fukazawa et al. (2000) found that the central iron excess is larger
(by 0.2-0.4 solar) in cooler clusters than in hotter ones (by 0.1-0.2
solar). This observation could provide circumstantial evidence for the
action of radio galaxies. In cooler clusters some time has passed
since the last epoch of radio activity and heating has occurred. The
hotter clusters, on the other hand, have recently experienced a period
of radio activity or are in the middle of such a period. If the radio
activity caused a redistribution of the metals, one would expect to
find a correlation between the central metal excesses and the radio
power of the cD galaxy. This conjecture could be checked
observationally. Alternatively, the observations by Fukazawa et
al. (2000) might merely reflect the fact that the hotter clusters may
have been disturbed recently by a merger which has mixed the metals
and reduced the central metallicity increment.

\acknowledgments 

Some of the computations reported here were performed using the UK
Astrophysical Fluids Facility (UKAFF). I thank Jim Pringle for helpful
discussions and the referee for useful comments.

\begin{figure}[htp]
\plottwo{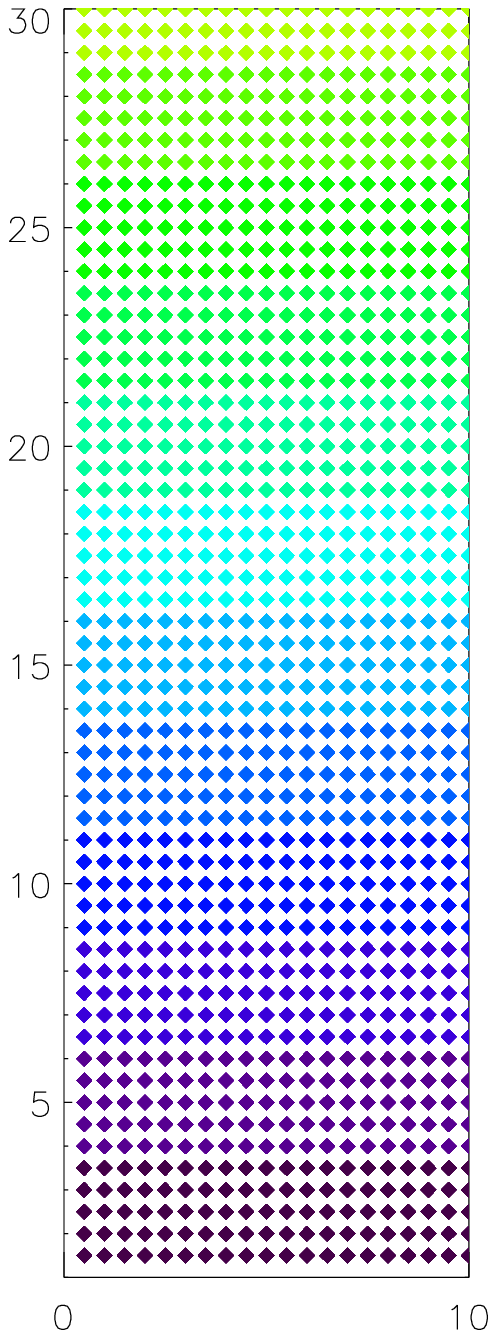}{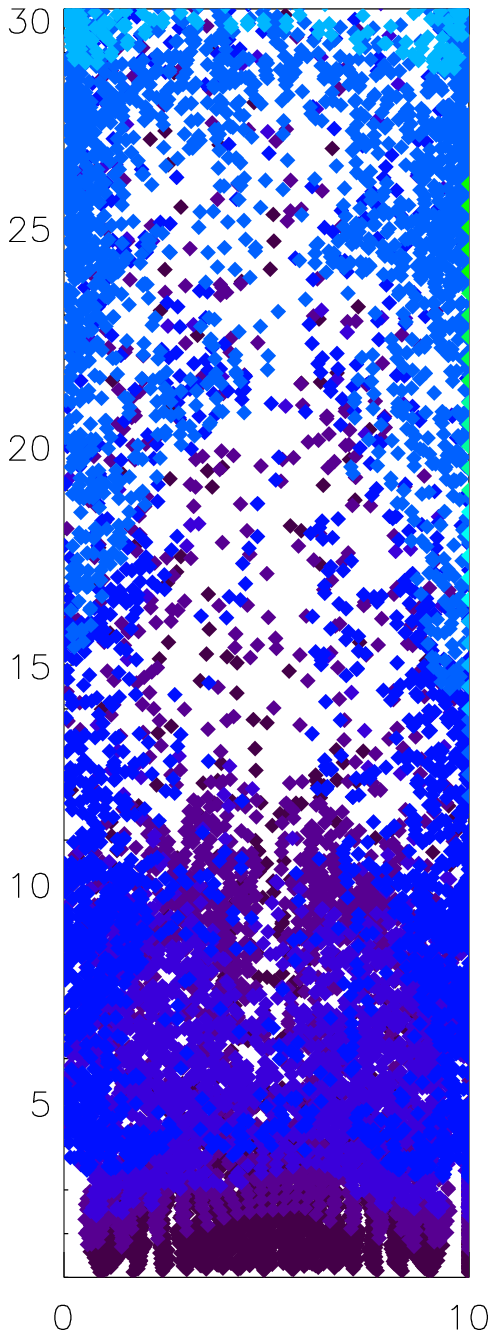}
\caption{Projected positions of the tracer particles at the start of
the simulation (left panel) and after a simulated time of 63 Myrs
(right panel) for run 2. The tracer particles are color coded according
to their initial positions as shown in the left panel. The center of
the cluster is at the bottom of the plot.}
\label{fig1}
\end{figure}

\begin{figure}[htp]
\plottwovert{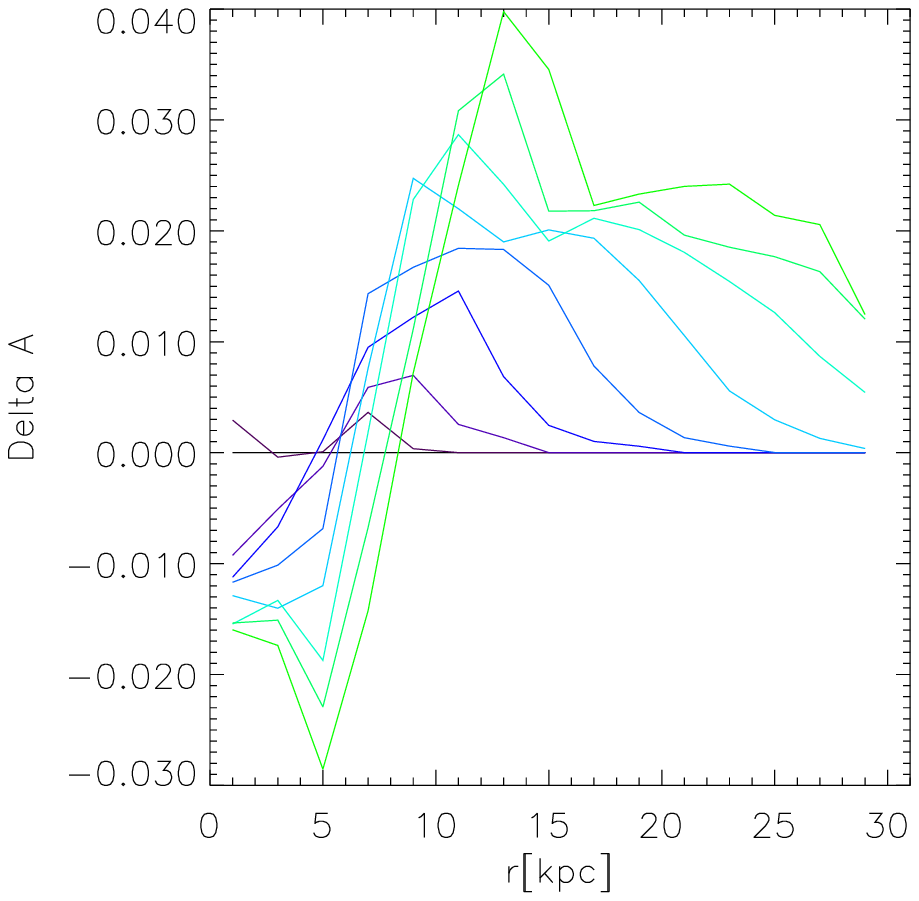}{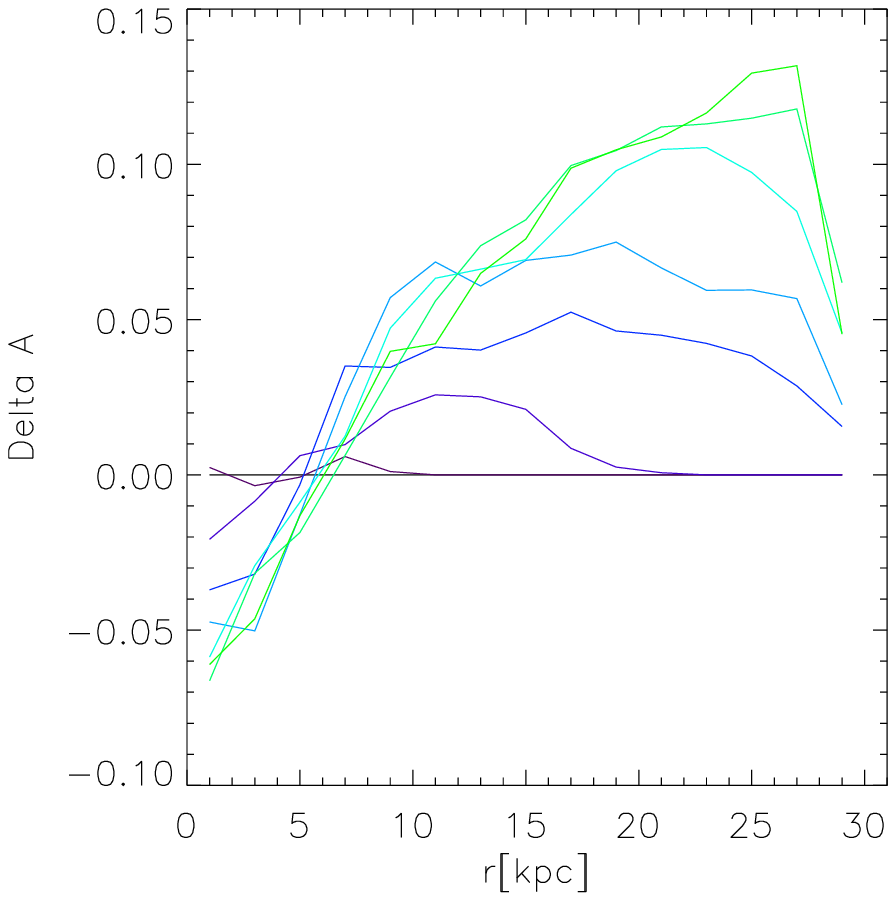}
\caption{Differences between the original iron abundance profile
(radius measured in kpc) and that at different times after the start
of the energy injection. The upper plot shows the results from run 1
and the lower ne from run 2. Curves in different colors correspond to
different times after the start of the simulation. The maximum
simulated time was 120 Myrs and the abundances have been averaged over
horizontal slices through the computational volume.}
\label{fig2}
\end{figure}

\begin{figure}[htp]
\plottwovert{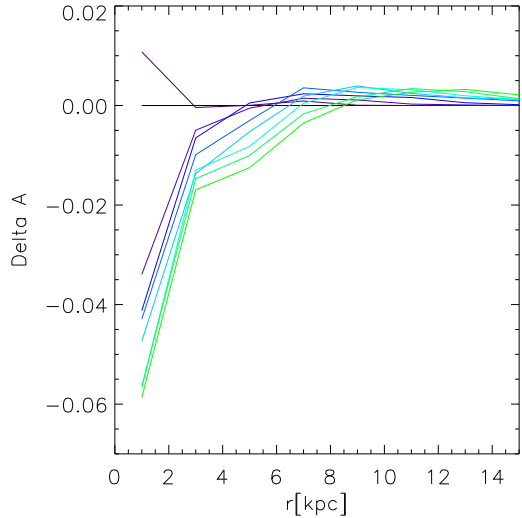}{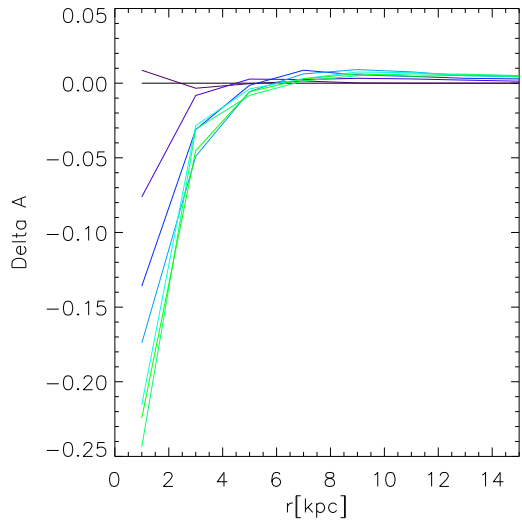}
\caption{Same as Fig.\ 2, only that now the abundances have been
averaged over semispherical shells. Again, the upper panel corresponds
to run 1 and the lower one to run 2. See text for details.}
\label{fig3}
\end{figure}

\label{lastpage}

\end{document}